\documentclass[fleqn,10pt]{SelfArx} % Document font size and equations flushed left

\definecolor{color1}{RGB}{0,0,90} % Color of the article title and sections
\definecolor{color2}{RGB}{0,20,20} % Color of the boxes behind the abstract and headings

\usepackage{hyperref} % Required for hyperlinks
\hypersetup{hidelinks,colorlinks,breaklinks=true,urlcolor=color2,citecolor=color1,linkcolor=color1,bookmarksopen=false,pdftitle={Title},pdfauthor={Author},urlcolor=blue}

\usepackage{graphicx}
\usepackage[square]{natbib}
\usepackage{amsmath}
\usepackage{multirow}
\usepackage{subfigure}
\usepackage{siunitx}

\newcommand{\n}[1]{\mathrm{#1}}

\JournalInfo{Published in Journal of Electronic Materials, Vol. 44 (8), 2869-2876, 2015} % Journal information
\Archive{\href{http://dx.doi.org/10.1007/s11664-015-3731-7}{DOI: 10.1007/s11664-015-3731-7}} % Additional notes (e.g. copyright, DOI, review/research article)

\PaperTitle{The universal influence of contact resistance on the efficiency of a thermoelectric generator} % Article title

\Authors{R. Bj\o{}rk} % Authors
\affiliation{\textit{Department of Energy Conversion and Storage, Technical University of Denmark - DTU, Frederiksborgvej 399, DK-4000 Roskilde, Denmark}} % Author affiliation
\affiliation{*\textbf{Corresponding author}: rabj@dtu.dk} % Corresponding author

 % Keywords - if you don't want any simply remove all the text between the curly brackets
\newcommand{\keywordname}{Keywords} % Defines the keywords heading name

\Abstract{The influence of electrical and thermal contact resistance on the efficiency of a segmented thermoelectric generator is investigated. We consider 12 different segmented $p$-legs and 12 different segmented $n$-legs, using 8 different $p$-type and 8 different $n$-type thermoelectric materials. For all systems a universal influence of both the electrical and thermal contact resistance is observed on the leg's efficiency, when the systems are analyzed in terms of the contribution of the contact resistance to the total resistance of the leg. The results are compared with the analytical model of Min and Rowe (1992). In order for the efficiency not to decrease more than 20\%, the contact electrical resistance should be less than 30\% of the total leg resistance for zero thermal contact resistance, while the thermal contact resistance should be less than 20\% for zero electrical contact resistance. The universal behavior also allowed the maximum tolerable contact resistance for a segmented system to be found, i.e. the resistance at which a leg of only the high temperature thermoelectric material has the same efficiency as the segmented leg with a contact resistance at the interface. If e.g. segmentation increases the efficiency by 30\% then an electrical contact resistance of 30\% or a thermal contact resistance of 20\% can be tolerated.}

\begin{document}

\flushbottom % Makes all text pages the same height

\maketitle % Print the title and abstract box

%\tableofcontents % Print the contents section

\thispagestyle{empty} % Removes page numbering from the first page

\section{Introduction}
Ensuring high efficiency for thermoelectric (TE) power generators is important, both scientifically and commercially. The inherent performance of a thermoelectric generator (TEG) is determined by the material properties, through the thermoelectric figure of merit, $ZT$, which is defined as
\begin{math}
ZT = \frac{\alpha^2T}{\rho\kappa}
\end{math}
where $\alpha$ is the Seebeck coefficient, $\rho$ is the resistivity, $\kappa$ is the thermal conductivity and $T$ is the absolute temperature. However, a number of secondary effects can strongly influence the efficiency of a TEG, most noticeably heat losses and thermal and electrical contact resistances. As heat losses can be avoided to some degree by proper insulation \cite{Ziolkowski_2010,Bjoerk_2014}, contact resistance is seen as the limiting factor for construction of high performing thermoelectric modules. This is especially so for segmented modules, where two thermoelectric materials are joined directly together to form a thermoelectric leg capable of spanning a large temperature range with high efficiency. For these segmented legs the contact resistance at the interface between the segmented materials is seen as detrimental to high efficiency.

That contact resistance causes a decrease in efficiency of experimental thermoelectric generators is well known \cite{ElGenk_2004,ElGenk_2006,Hung_2014} and an increase in total leg resistance by a factor of 2-3 by adding electrodes to a single Mg$_2$Si leg have been observed \cite{Sakamoto_2012}. Several studies discuss the fabrication of low electrical resistance contacts for thermoelectrics, using e.g. metal contacts \cite{DAngelo_2007}, titanium disilicide (TiSi$_2$) \cite{Assion_2013}, transition-metal silicides \cite{Sakamoto_2012}, silver-based alloys \cite{Gan_2013}, antimony (Sb) \cite{Li_2012}, Ti foil \cite{Zhao_2012} and Ag and Cu \cite{Zybala_2010}. Regarding the value of the contact resistance, a specific electrical contact resistance of $\sim{}10^{-5}$ $\mathrm{\Omega}$ cm$^2$ can be realized experimentally \cite{DAngelo_2007,Li_2012}. Experimentally, the contact resistance at the interface of thermoelectric materials can arise either due to surface roughness or it can be directly related to the interfaces of the materials formed during sintering \cite{Zhao_2012}.

The actual influence of a contact resistance, either electrical or thermal, on the efficiency of a segmented thermoelectric generator is not known in detail. Experimentally, the contact resistance is known to reduce the efficiency of a segmented module \cite{ElGenk_2003,DAngelo_2011}. For the case of a thermal contact resistance, the resistance will lower the temperature span across the individual segments of the leg. This will cause a decrease in efficiency, but this decrease will depend on the material properties and cannot in general be predicted analytically. Even for constant material properties, the generator efficiency is a complicated function of both hot and cold side temperatures, $T_\n{h}$ and $T_\n{c}$ respectively, as well as temperature span, $\Delta{}T$. In this case the efficiency, $\eta$, is given by \cite{TE_Handbook_Ch9}
\begin{eqnarray} \label{Eq.Eta_TE_hand}
\eta = \frac{\Delta{}T}{T_\n{h}}\frac{\sqrt{1+Z\bar{T}}-1}{\sqrt{1+Z\bar{T}}+T_\n{c}/T_\n{h}}
\end{eqnarray}
for constant material properties, where $\bar{T}$ is the mean temperature. For material parameters that are a function of temperature, the drop in efficiency as function of temperature span cannot be predicted analytically.

An expression for the efficiency of a thermoelectric leg, in the presence of both an electrical and thermal contact resistance have been derived by Min and Rowe \cite{Min_1992,Rowe_1996,TE_Handbook_Ch9}, assuming constant material properties as function of temperature. The efficiency is given as
\begin{eqnarray}\label{Eq.Min_Rowe}
%\eta = \frac{\frac{\Delta{}T}{T_h}}{\left(1+2\frac{R_\n{c,T}}{R_\n{T}}\right)^2\left(2-\frac{1}{2}\frac{\Delta{}T}{T_h}+\frac{4}{z T_h}\left(\frac{1+2\frac{R_\n{c,e}}{R_\n{e}}}{1+2\frac{R_\n{c,T}}{R_\n{T}}}\right)\right)}
\eta = \frac{\Delta{}T}{T_h}\frac{1}{\left(1+2\frac{R_\n{c,T}}{R_\n{l,T}}\right)^2\left(2-\frac{1}{2}\frac{\Delta{}T}{T_h}+\frac{4}{z T_h}\frac{1+2R_\n{c,e}/R_\n{l,e}}{1+2R_\n{c,T}/R_\n{l,T}}\right)}
\end{eqnarray}
where $R$ is the resistance, the subscript $c$ or $l$ denotes the contact or leg resistence, respectively, while $e$ or $T$ denotes the electrical or thermal resistance, respectively. Thus, e.g. $R_\n{c,e}$ is the total electrical contact resistance in Ohms\footnote{In the derivation of Eq. (\ref{Eq.Min_Rowe}), we have assumed that there is a typo in the expression given in Ref. \cite{TE_Handbook_Ch9}, Eq. (11.4). A factor of $l_c$ is missing in the last parenthesis in the denominator, i.e. the equation should be $(l+nl_c)/(l+2rl_c)$.}. This expression does not consider material properties that are a function of temperature, nor does it in principle consider segmented legs.

Numerically, only the influence of electrical contact resistance has been examined \cite{Pettes_2007, Ebling_2010,Reddy_2014}. These studies all show a tangent hyperbolic-like function behavior of efficiency as function of specific contact resistance, however all studies have only consider a single material, namely bismuth telluride.

Here we will consider the influence of both an electrical and thermal contact resistance on a segmented thermoelectric module, using a numerical model. We will consider a large number of different thermoelectric materials, and differently segmented legs, in order to elucidate if a general trend exists on the influence of contact resistance on the efficiency of a thermoelectric generator.

\section{TE materials}
\begin{figure}[!t]
  \centering
  \includegraphics[width=1\columnwidth]{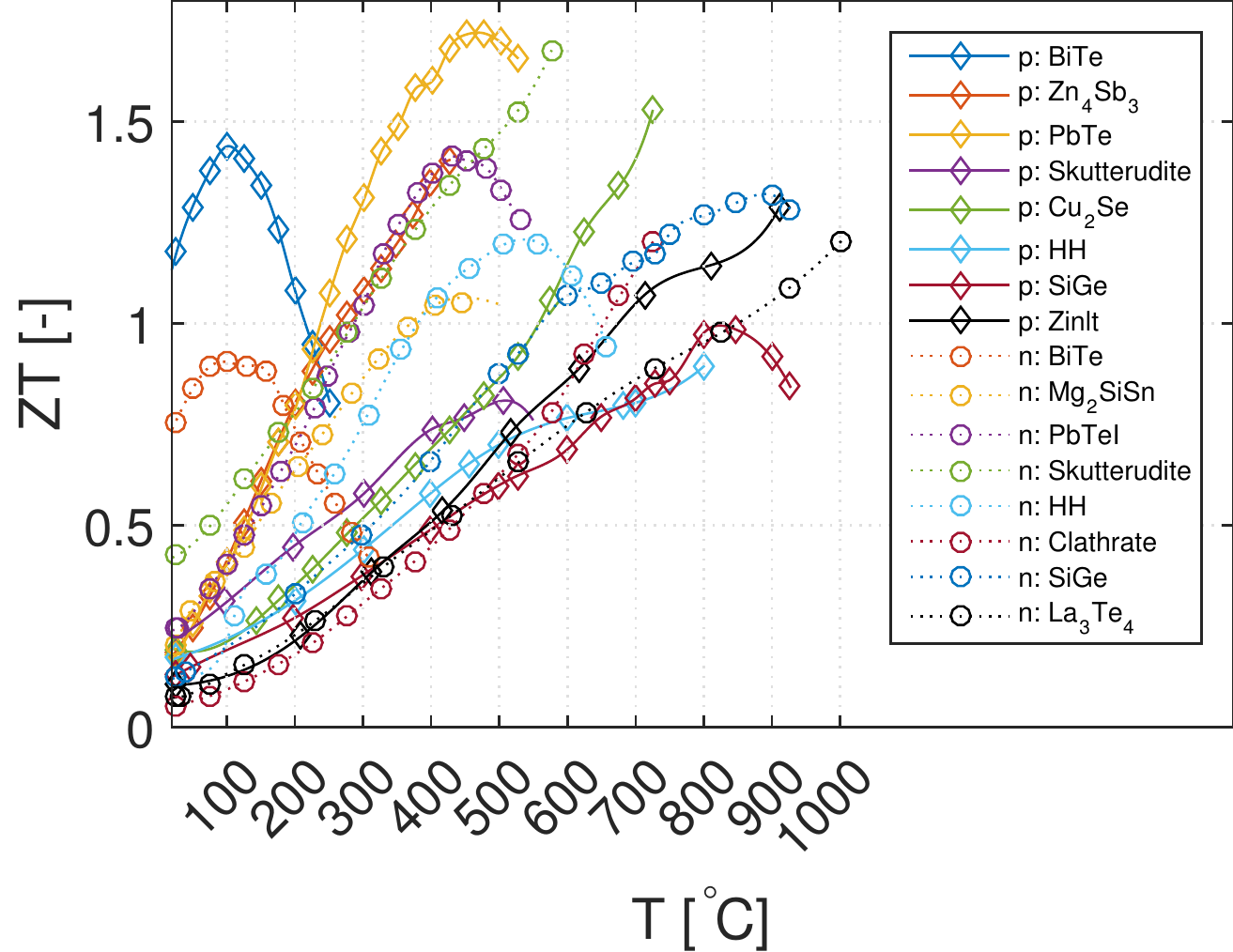}
  \caption{The $ZT$ value of the 16 different thermoelectric materials considered as function of temperature. Only $ZT$ is shown for clarity, but all relevant material parameters have been obtained.}
  \label{Fig_Mat_zT}
\end{figure}

For this study of the influence of contact resistance we consider 8 $p$-type and 8 $n$-type TE materials. The specific $p$-type materials considered are BiSbTe \cite{Ma_2008}, NdFe$_{3.5}$Co$_{0.5}$Sb$_{12}$ (Skutterudite) \cite{Muto_2013}, Yb$_{14}$Mn$_{0.2}$Al$_{0.8}$Sb$_{11}$ (Zinlt) \cite{Toberer_2008}\\, Zr$_{0.5}$Hf$_{0.5}$CoSb$_{0.8}$Sn$_{0.2}$ (Half-Heusler, HH) \cite{Yan_2010}, PbTe \cite{Pei_2011}, Zn$_4$Sb$_3$ \cite{Chitroub_2008}, Cu$_2$Se \cite{Liu_2012} and SiGe \cite{Joshi_2008}. The specific $n$-type materials considered are BiTe \cite{Kim_2012},\\ Ti$_{0.5}$Zr$_{0.25}$Hf$_{0.25}$NiSn$_{0.998}$M$_{0.002}$ (Half-Heusler, HH) \cite{Schwall_2013},\\ Ba$_{8}$Ni$_{0.31}$Zn$_{0.52}$Ga$_{13.06}$Ge$_{32.2}$ (Clathrate) \cite{Shi_2010}, \\ Mg$_{2}$Si$_{0.3925}$Sn$_{0.6}$Sb$_{0.0075}$ \cite{Zhang_2008}, PbTe$_{0.9988}$I$_{0.0012}$ \cite{Lalonde_2011}, \\ Ba$_{0.08}$La$_{0.05}$Yb$_{0.04}$Co$_{4}$Sb$_{12}$ (Skutterudite) \cite{Shi_2011},\\ La$_{3}$Te$_{4}$ \cite{May_2010} and SiGe \cite{Wang_2008}. All materials have temperature dependent experimentally measured properties, and the calculated $ZT$ values for the different materials are shown in Fig. \ref{Fig_Mat_zT}. Some, though not all, of the materials are similar to those considered by Ref. \cite{Ngan_2014}. All material parameters, as function of temperature, have been obtained from the cited references and are used in the following calculations, but only $ZT$ is shown for clarity. The remaining material properties are available from the author upon request, or are of course available from the cited references.

In order to investigate the influence of contact resistance on a segmented TE leg, we examine 12 combinations of two $p$-type materials and 12 combinations of two $n$-type materials. While there is a total of at least 28 different combinations for both a $p$- and $n$-type leg for the materials above, not all combinations will result in a leg with an increased efficiency, nor is not necessary to examine all possible combinations in order to draw general conclusions on the influence of contact resistance.

\begin{figure}[!t]
  \centering
  \includegraphics[width=1\columnwidth]{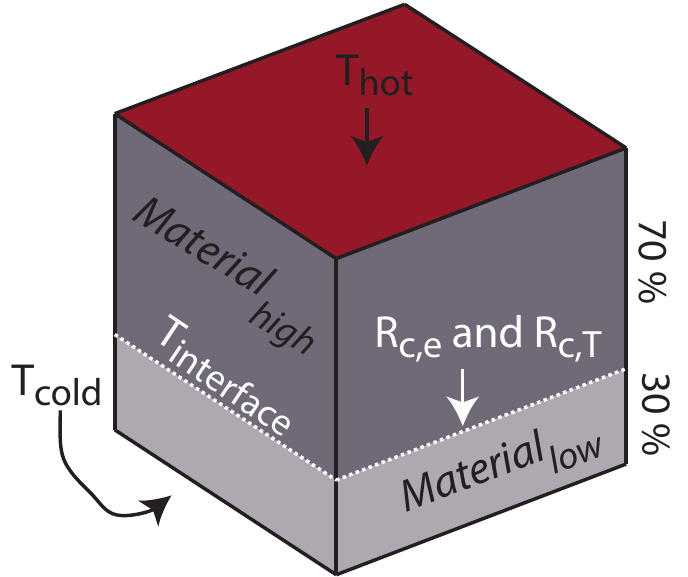}
  \caption{The setup considered consists of two TE materials, segmented together to form a TE-leg. The subscript ``low'' indicates the material at low temperature while ``high'' indicates the material at high temperature. The contact resistance is added at the interface between the TE materials. The optimal volume ratio of the low and high materials are determined for each of the segmented systems considered. In this illustration their sizes are 30\% and 70\%, respectively.}
  \label{Fig_Illustration_single_leg}
\end{figure}

A numerical Comsol model, which includes all relevant thermoelectric phenomena, is used to calculate the efficiency of a segmented TE leg \cite{Bjoerk_2014}. The model fully accounts for all material parameters, and all as function of temperature. No heat loss is assumed in the current work and we only consider the efficiency of single leg of either $p$- or $n$-type material, as shown in Fig. \ref{Fig_Illustration_single_leg}. We consider the efficiency, $\eta$, defined as
\begin{eqnarray}\label{Eq.Efficiency_def}
\eta{}=\frac{P}{Q_\n{in}}
\end{eqnarray}
where $P$ is the electrical power produced by the leg at the optimal load resistance, and $Q_\n{in}$ is the heat flowing into the leg.

In order to determine the influence of contact resistance on a segmented leg, we initially calculated the efficiency without any contact resistance for the 12 different $p$-type and 12 $n$-type systems given in Table \ref{Table.p} and \ref{Table.n}. The external resistance and the volume ratio of the low- to high temperature TE materials was varied to determine the optimal segmented leg. The hot side temperature was selected based on the peak $ZT$ temperature for the hot side material. A maximum temperature, $T_\n{max}$, which is the highest temperature at which the material properties were reported exist for the low temperature material. The cold side temperature was kept constant at 20 $^\circ{}$C. The geometry of the optimal segmented legs are given in Table \ref{Table.p} and \ref{Table.n}, respectively. In these tables it can be seen that segmenting a TE leg in general increases the efficiency of the total system. This is seen as the maximum efficiency of the segmented leg, $\eta_\n{max}$, is larger than the efficiency of a single leg consisting of only the high temperature material, $\eta_\n{max, no-seg}$. Also, in all cases except one, the segmented interface temperature, $T_\n{interface}$, is equal to the maximum temperature of low temperature material, $T_\n{max}$, as is expected as the lower temperature material generally have a higher $ZT$ value \cite{Saber_2002}. Of course, in order to choose which system to optimally segment one would have to study the compatibility factors of the given materials \cite{Ursell_2002}. The ratio of the mean compatibility factors of the two segmented materials, $\bar{s}$, have been calculated and are given in Table \ref{Table.p} and \ref{Table.n}. In general this factor has to be within about a factor of 2 \cite{TE_Handbook_Ch9}, and as can be seen this is the case for most segmented legs considered. However, we also consider segmented legs with a larger compatibility factor ratio, in order to investigate how these are influenced by contact resistance as well.

%--- p-type
\begin{table*}
\begin{tabular}{c c c c c | c S[table-format=3.2] c | S[table-format=3.2] | c}
%Mat$_\n{low}$ & Mat$_\n{high}$ & $T_\n{hot}$ & $T_\n{max}$ for mat$_\n{low}$ & Optimal $T_\n{interface}$ & Amount of mat$_\n{low}$ & $\eta_\n{max}$ \\ %\hline
Mat$_\n{low}$ & Mat$_\n{high}$ & $T_\n{hot}$ & $T_\n{max}$ & $\bar{s}$ & Optimal & {Amount}  & $\eta_\n{max}$ & {$\eta_\n{max}$ of} & Gain by \\ %\hline
& & & for mat$_\n{low}$ & ratio & $T_\n{interface}$ & {of mat$_\n{low}$} & & {no-seg. mat$_\n{high}$ only} & segmentation\\
 & & [$^\circ{}$C] & [$^\circ{}$C] & & [$^\circ{}$C] & {[\%]} & [\%] & {[\%]} & [pp.] \\ \hline
BiTe & Cu2Se         & 700 & 200 & 3.0 & 200 & 27.4 & 14.30 & 11.18 & 3.12 \\
BiTe & HH            & 700 & 200 & 3.7 & 200 &  9.8 & 12.61 &  9.87 & 2.74 \\
BiTe & PbTe          & 500 & 200 & 1.9 & 200 & 31.4 & 16.50 & 13.23 & 3.27 \\
BiTe & Skutterudite  & 500 & 200 & 2.1 & 200 & 21.4 & 13.24 &  8.62 & 4.62 \\
BiTe & Zn4Sb3        & 400 & 200 & 1.6 & 200 & 60.0 & 14.14 &  9.87 & 4.27 \\
BiTe & SiGe          & 900 & 200 & 3.7 & 200 &  9.3 & 13.86 & 11.20 & 2.66 \\
BiTe & Zinlt         & 900 & 200 & 3.6 & 200 & 28.6 & 14.98 & 11.84 & 3.14 \\
Cu2Se & SiGe         & 900 & 700 & 1.6 & 700 & 53.4 & 13.30 & 11.20 & 2.10 \\
Skutterudite & SiGe  & 900 & 500 & 2.0 & 500 & 48.9 & 11.97 & 11.20 & 0.77 \\
PbTe & Zinlt         & 900 & 500 & 2.0 & 500 & 73.1 & 17.29 & 11.84 & 5.45 \\
Zn4Sb3 & Zinlt       & 900 & 400 & 2.2 & 400 & 43.3 & 14.89 & 11.84 & 3.05 \\
Skutterudite & Zinlt & 900 & 500 & 1.9 & 493 & 79.4 & 13.14 & 11.84 & 1.30 \\
\end{tabular}
\caption{The segmented $p$-type TE legs considered. The subscript ``low'' and ``high'' refer to the materials placed at high and low temperature, respectively. The $\bar{s}$-ratio is the ratio of the mean compatibility factor of the two materials. The last column gives the increase in efficiency in percentage points (pp.) gained by segmentation, i.e. $\eta_\n{max}$-$\eta_\n{max, no-seg}$. The cold side temperature was kept constant at 20 $^\circ{}$C.} \label{Table.p}
\end{table*}

%--- n-type
\begin{table*}
\begin{tabular}{c c c c c | c S[table-format=3.2] c | S[table-format=3.2] | c}
Mat$_\n{low}$ & Mat$_\n{high}$ & $T_\n{hot}$ & $T_\n{max}$ & $\bar{s}$ & Optimal & {Amount}  & $\eta_\n{max}$ & {$\eta_\n{max}$ of} & Gain by \\ %\hline
& & & for mat$_\n{low}$ & ratio & $T_\n{interface}$ & {of mat$_\n{low}$} & & {no-seg. mat$_\n{high}$ only} & segmentation\\
 & & [$^\circ{}$C] & [$^\circ{}$C] & &  [$^\circ{}$C] & {[\%]} & [\%] & {[\%]} & [pp.] \\ \hline
BiTe & Clathrate        & 700  & 200 & 2.7 & 200 & 20.6 & 11.35 & 7.97  & 3.38 \\
BiTe & La3Te4           & 1000 & 200 & 4.0 & 200 & 23.1 & 13.74 & 12.24 & 1.50 \\
BiTe & Mg2SiSn          & 500  & 200 & 2.3 & 200 & 20.1 & 11.97 & 10.92 & 1.02 \\
BiTe & PbTeI            & 500  & 200 & 1.5 & 200 & 29.8 & 14.05 & 11.88 & 2.17 \\
BiTe & Skutterudite     & 550  & 200 & 1.5 & 200 & 16.1 & 15.40 & 14.18 & 1.22 \\
BiTe & HH               & 600  & 200 & 2.3 & 200 & 14.9 & 13.67 & 10.72 & 2.95 \\
BiTe & SiGe             & 900  & 200 & 3.1 & 200 & 9.0  & 15.48 & 14.00 & 1.48 \\
Skutterudite & La3Te4   & 1000 & 550 & 3.1 & 550 & 78.6 & 15.04 & 12.24 & 2.80 \\
Mg2SiSn & La3Te4        & 1000 & 500 & 1.9 & 500 & 74.6 & 15.80 & 12.24 & 3.56 \\
Skutterudite & SiGe     & 900  & 550 & 2.3 & 550 & 55.4 & 16.53 & 14.00 & 2.53 \\
Mg2SiSn & SiGe          & 900  & 500 & 1.4 & 500 & 49.9 & 16.87 & 14.00 & 2.87 \\
PbTe-I & SiGe           & 900  & 500 & 2.3 & 500 & 40.4 & 15.89 & 14.00 & 1.89 \\
\end{tabular}
\caption{The segmented $n$-type TE legs considered. The cold side temperature was kept constant at 20 $^\circ{}$C} \label{Table.n}
\end{table*}

\section{Influence of contact resistance}
Having determined the ideal geometry of the different segmented legs considered, a thermal and electrical contact resistance was introduced at the interface between the two TE materials and varied systematically while the efficiency was computed. Always, for each system the optimal load resistance was determined, i.e. the load resistance resulting in the highest efficiency. The varied values of the specific electrical contact resistance was $R_\n{c,e} = [10^{-10},\; 5*10^{-9},\;10^{-9},\;5*10^{-8},\;10^{-8},\; 5*10^{-7},\;10^{-7},\; 5*10^{-6},\; 10^{-6},\; 5*10^{-5}]$ $\mathrm{\Omega}$ m$^2$, while the varied values of the specific thermal contact resistance was $R_\n{c,T} = [10^{-6},\; 5*10^{-5},\; 10^{-5},\; 5*10^{-4},\; 10^{-4},\; 5*10^{-3},\; 10^{-3},\; 5*10^{-2},\; 10^{-2},\; 5*10^{-1}]$ m$^2$KW$^{-1}$, both varied independently.

Without contact resistance the efficiency does not depend on the length or the cross-sectional area of the leg. When a specific contact resistance is added, there is still no dependence on the cross-sectional area, but the efficiency will now depend on the length of the leg. In order to remove this dependence on geometry, all resistance are characterized as the total, and not the specific, resistance. Thus the results must be understood in terms of the fraction of total contact resistance to total overall resistance, $R_\n{c}/R_\n{total}$, which is a variable that does not depend on the length of the leg. Here the total overall resistance is simply $R_\n{total}=R_\n{c}+R_\n{leg}$. Therefore all results presented here are valid for all leg geometries.

\begin{figure}[!t]
  \centering
  \includegraphics[width=1\columnwidth]{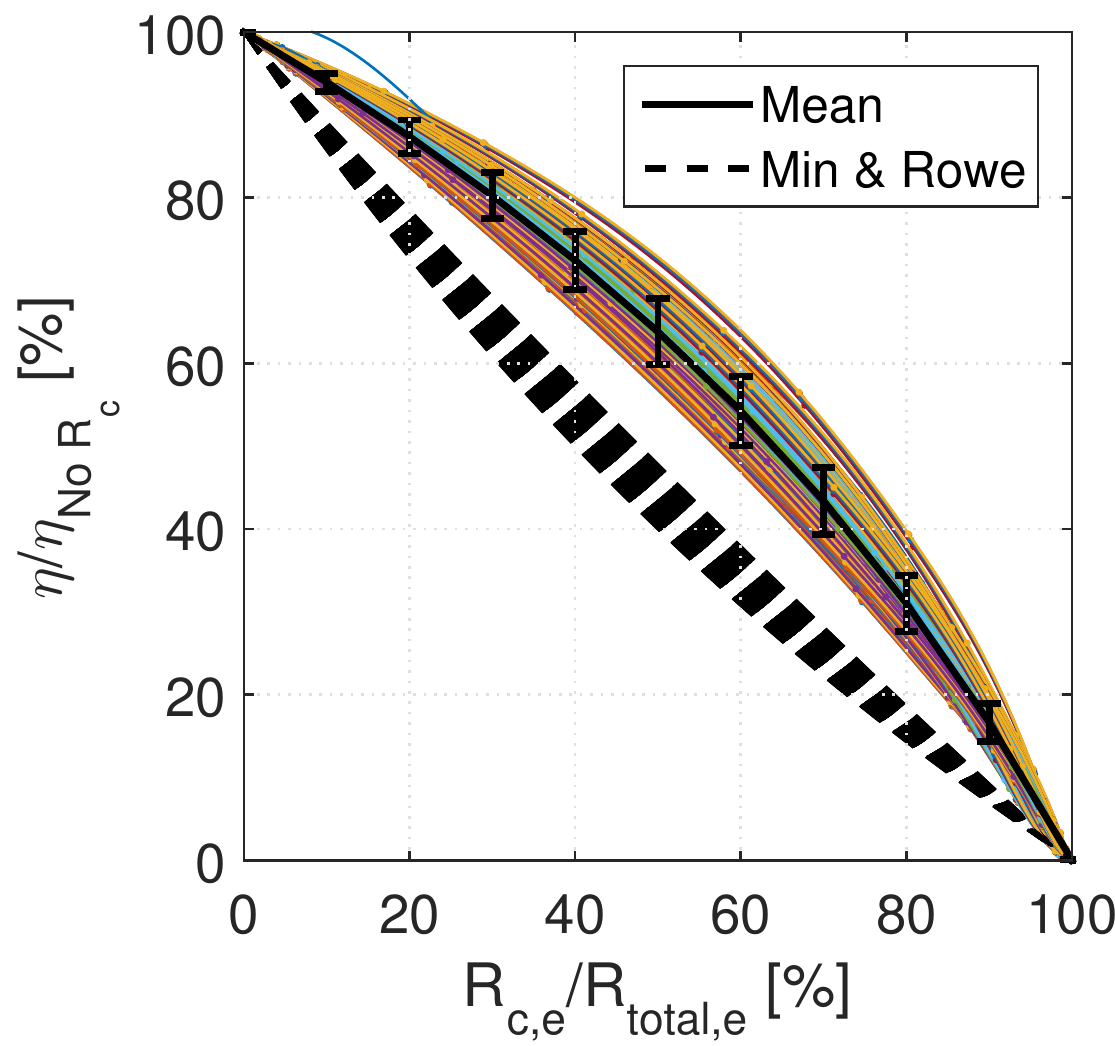}
  \caption{The decrease in efficiency, $\eta$, as function of the fraction of total electrical contact resistance to total resistance, for all systems and all thermal contact resistances considered. A total of 240 curves are shown, e.g. for 24 segmented legs with 10 different thermal contact resistances at the interface. The curves are spline interpolated for the 10 data points for the specific electrical contact resistance. The model of Min \& Rowe \cite{Min_1992} is also shown.}
  \label{Fig_Rc_Eff}
\end{figure}

Shown in Fig. \ref{Fig_Rc_Eff} is the decrease in efficiency as function of electrical contact resistance for all thermal contact resistances considered and for all $p$- and $n$-segmented legs. The electrical contact resistance is normalized in terms of the total electrical resistance of the system, i.e. the sum of the resistance of the contact and the resistance of the TE materials. As can be seen from the figure the general behavior of the curves are the same, indicating a universal behavioral influence of the electrical contact resistance on the efficiency of the TEG. The mean of all 240 systems considered is also shown, with the error bars indicating the standard deviation. The predicted decrease in performance using Eq. (\ref{Eq.Min_Rowe}) is also shown in the figure, assuming a $ZT$ value that produces an efficiency equal to that of the leg without contact resistance. It is seen that the analytical expression overestimates the decrease in efficiency. The reason for this is partly the assumption of constant material properties, as well as the segmented leg geometry, i.e. the internal location of the contact resistance.

\begin{figure}[!t]
  \centering
  \includegraphics[width=1\columnwidth]{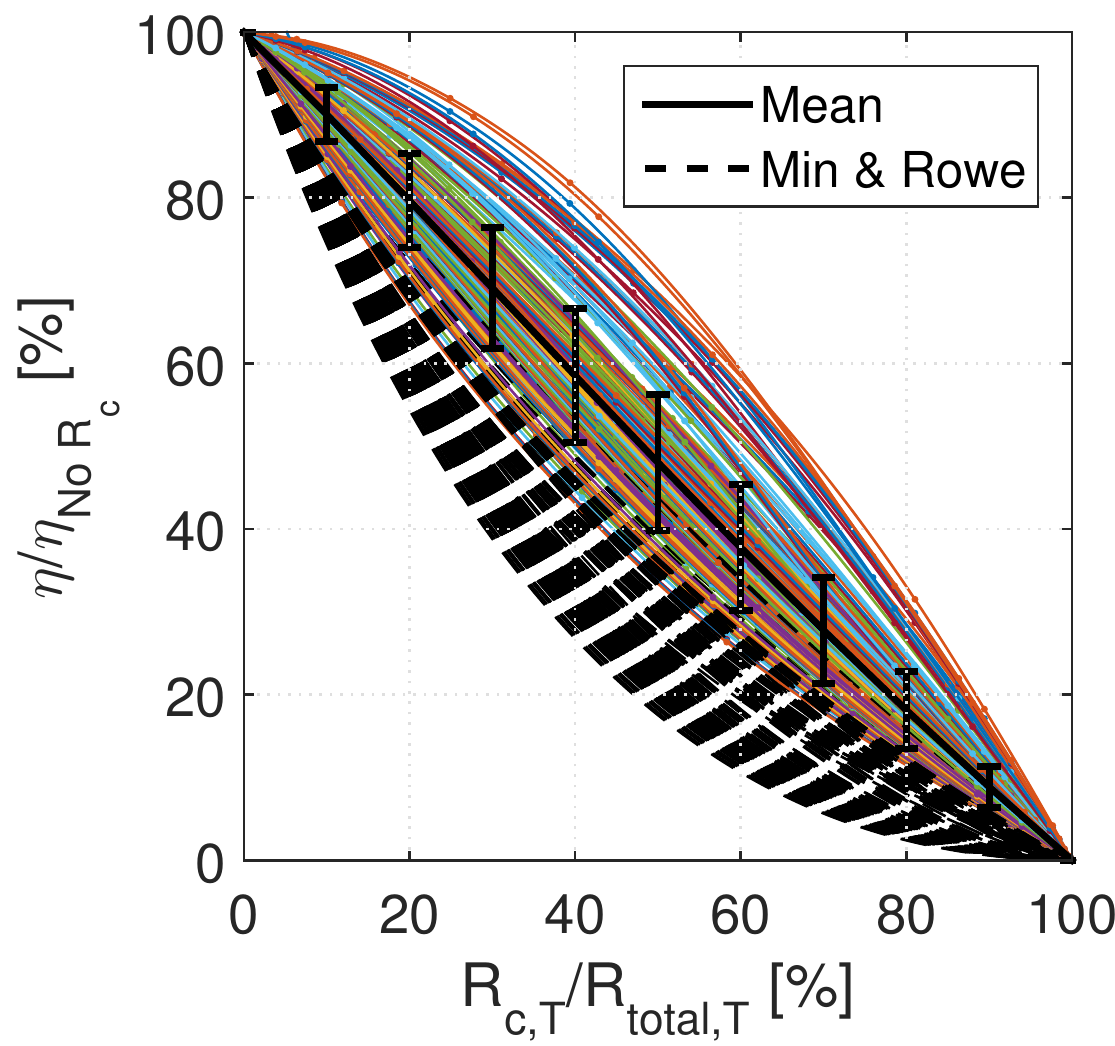}
  \caption{The decrease in efficiency, $\eta$, as function of the total thermal contact resistance, for all systems and all electrical contact resistances considered. A total of 240 curves are shown, e.g. for 24 segmented legs with 10 different electrical contact resistances at the interface. The curves are spline interpolated for the 10 data points for the specific thermal contact resistance. The model of Min \& Rowe \cite{Min_1992} is also shown.}
  \label{Fig_Rt_Eff}
\end{figure}

Similarly as Fig. \ref{Fig_Rc_Eff}, the decrease in efficiency as function of thermal contact resistance for all electrical contact resistances considered is shown in Fig. \ref{Fig_Rt_Eff}. The thermal contact resistance is normalized in terms of the total thermal resistance of the system. Similarly with the electrical contact resistance, the general behavior of the curves are the same for all legs considered, indicating a universal behavior. However, the spread in the curves are larger than in Fig. \ref{Fig_Rc_Eff}. This is because a thermal contact resistance changes the temperature throughout the leg, which influences the material properties and thus the generated power, while this is not the case for an electrical contact resistance. The mean of all 240 systems considered are also shown, with the error bars indicating the standard deviation. Again the analytical expression in most cases overestimates the decrease in efficiency. As the temperature is changed internally in the leg, due to the thermal contact resistance, the material properties change, which is not captured by the analytical model.

The tangent hyperbolic-like function behavior for efficiency as function of specific contact resistance seen by Refs. \cite{Pettes_2007, Ebling_2010,Reddy_2014} is identical to the curves shown in Fig. \ref{Fig_Rc_Eff}, expect that the variable have been changed to reflect the fraction of contact resistance to the total resistance, instead of specific contact resistance. The curves of the decrease in efficiency as function of thermal contact resistance are also tangent hyperbolic-like when plotted as function of specific contact resistance.

\begin{figure}[!t]
  \centering
  \includegraphics[width=1\columnwidth]{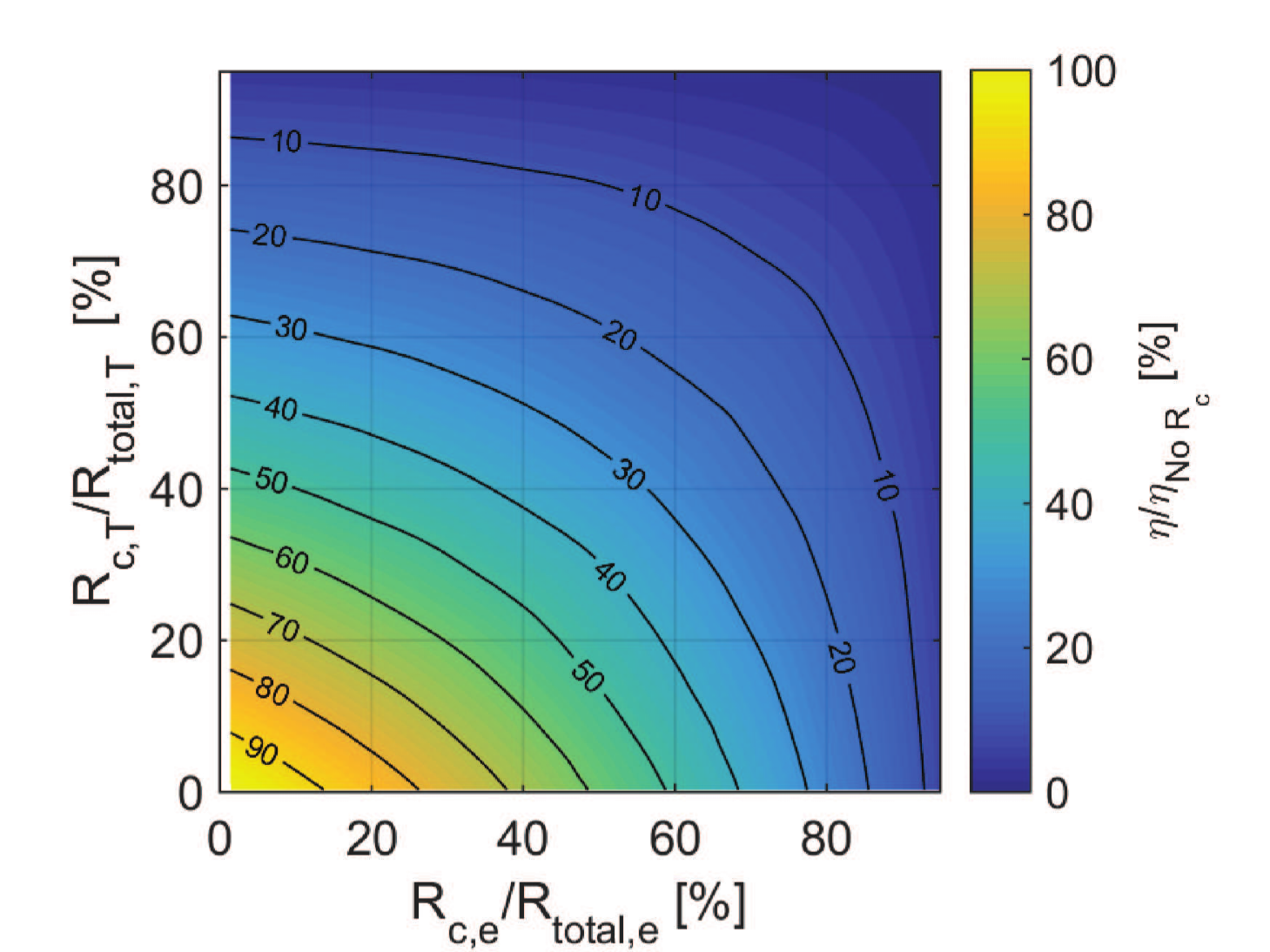}
  \caption{The average decrease in efficiency, $\eta$, as function of the total amount of electrical and thermal contact resistance. The average at each point is over all 24 segmented legs considered.}
  \label{Fig_Eff_surf_Rc_Rt}
\end{figure}

\begin{figure}[!t]
  \centering
  \includegraphics[width=1\columnwidth]{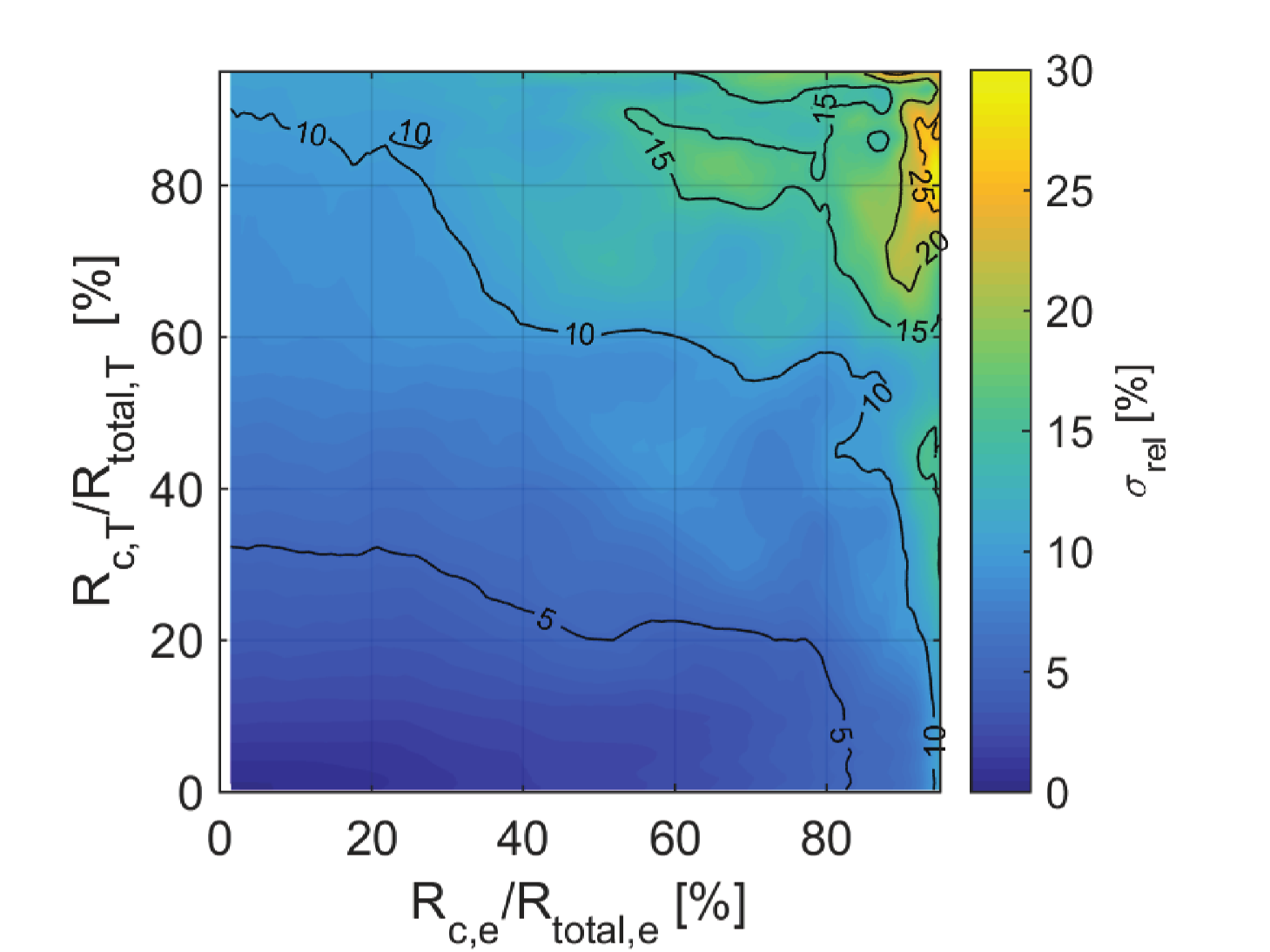}
  \caption{The relative standard deviation, $\sigma_\n{rel}$, of the efficiency as function of the total amount of electrical and thermal contact resistance, i.e. the efficiency shown in Fig. \ref{Fig_Eff_surf_Rc_Rt}.}
  \label{Fig_Sigma_surf_Rc_Rt}
\end{figure}

The spreading of the curves in Figs. \ref{Fig_Rc_Eff} and \ref{Fig_Rt_Eff}, for a given contact resistance is due to the dependence of efficiency on $ZT$ and hot side temperature, as given in Eq. (\ref{Eq.Eta_TE_hand}), as well as on the variation of the ``other'' contact resistance, i.e. thermal contact resistance for the case of electric contact resistance, and vice versa. Thus, if both a electrical and thermal contact resistance is present in the system, there will be a decrease in efficiency caused by both. This dual effect can be eliminated by considering the decrease in efficiency as function of both types of contact resistance. This decrease in efficiency in shown in Fig. \ref{Fig_Eff_surf_Rc_Rt}, which shows the mean normalized efficiency of all systems considered as function of the normalized electrical and thermal contact resistances. The corresponding relative standard deviation for the mean values at each point is shown in Fig. \ref{Fig_Sigma_surf_Rc_Rt}. As can be seen from the latter figure, the relative standard deviation for the normalized electrical and thermal contact resistances is less than 10\% for contact resistances less than 80\%. This indicates a universal behavior, where the decrease in efficiency for all thermoelectric system can be estimated fairly accurately using Fig. \ref{Fig_Eff_surf_Rc_Rt}, as long as the resistances are known. The variation indicated by Fig. \ref{Fig_Sigma_surf_Rc_Rt} is caused by the varying material properties and temperature spans on the considered materials, i.e the variation in efficiency that can be predicted using Eq. (\ref{Eq.Eta_TE_hand}). The key observation is that this variation is small, i.e. that there is a universal behavior, even considering the variety of the different material systems considered.

\section{The benefit of segmentation}
\begin{figure}[!t]
  \centering
  \includegraphics[width=1\columnwidth]{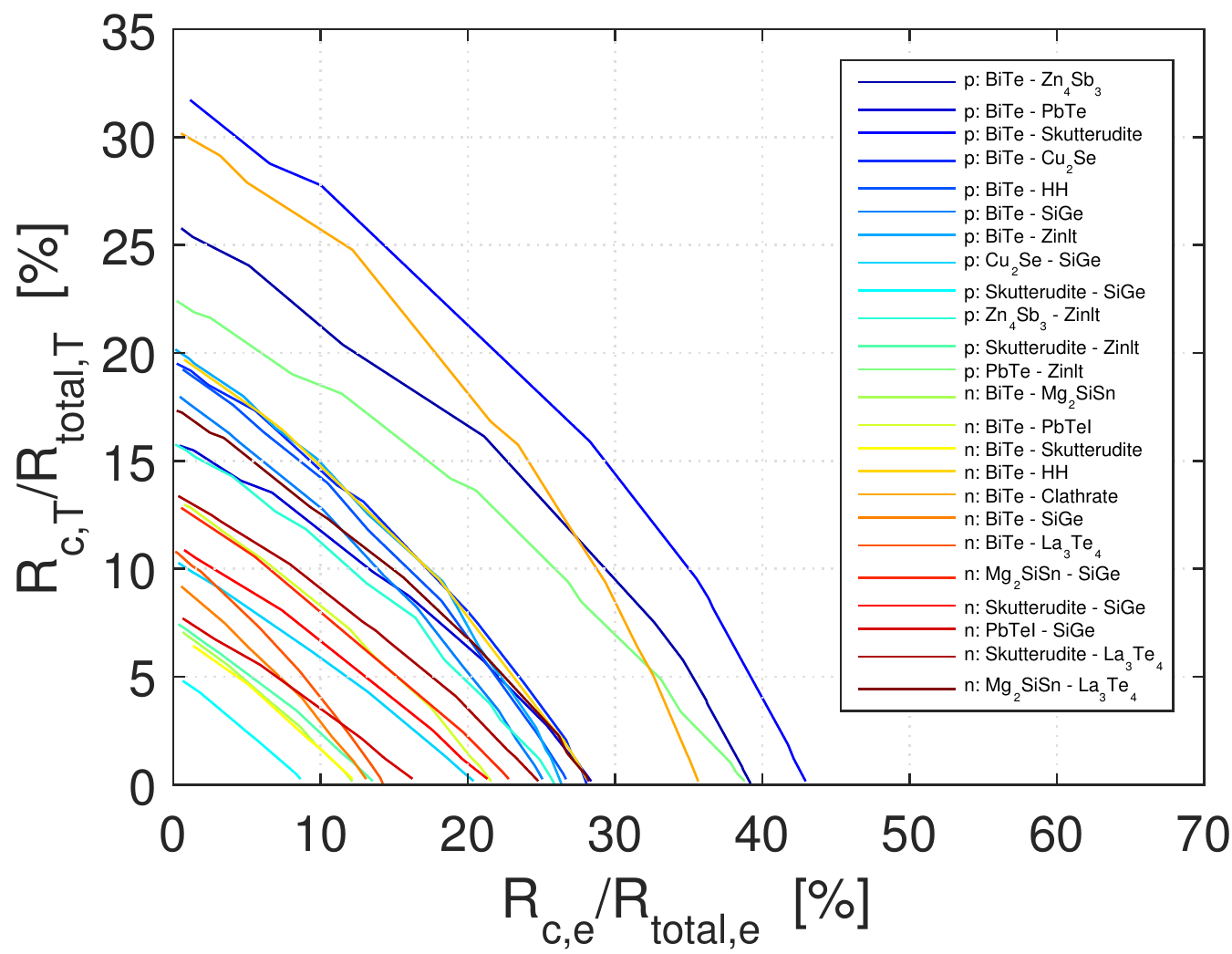}
  \caption{Contours of the maximum tolerable combined electrical and thermal contact resistance for a segmented TE leg. At the contour, efficiency of a single leg of only high temperature material is equal to that of a segmented leg with contact resistance at the segmentation interface.}
  \label{Fig_Maximum_eff_seg_Rc}
\end{figure}

The results presented in Fig. \ref{Fig_Eff_surf_Rc_Rt} can also be used to estimate whether segmenting two given thermoelectric materials is worth considering or not. If the joining between the materials cannot be made such that drop in efficiency caused by the contact resistance at the interface is lower than the gain in efficiency due to segmentation, then the materials should not be segmented. As the efficiency of a single leg of only high temperature material is known for each system (see Table \ref{Table.p} and \ref{Table.n}), the point in the parameter space where the above statement is true can be determined in Fig. \ref{Fig_Eff_surf_Rc_Rt} as function of both thermal and electrical contact resistance. At these points, it will be equally efficient to have a single leg of only high temperature material as compared with a segmented leg with contact resistance at the segmentation interface. These points are shown in Fig. \ref{Fig_Maximum_eff_seg_Rc} for all systems considered.

\begin{figure}[!t]
  \centering
  \includegraphics[width=1\columnwidth]{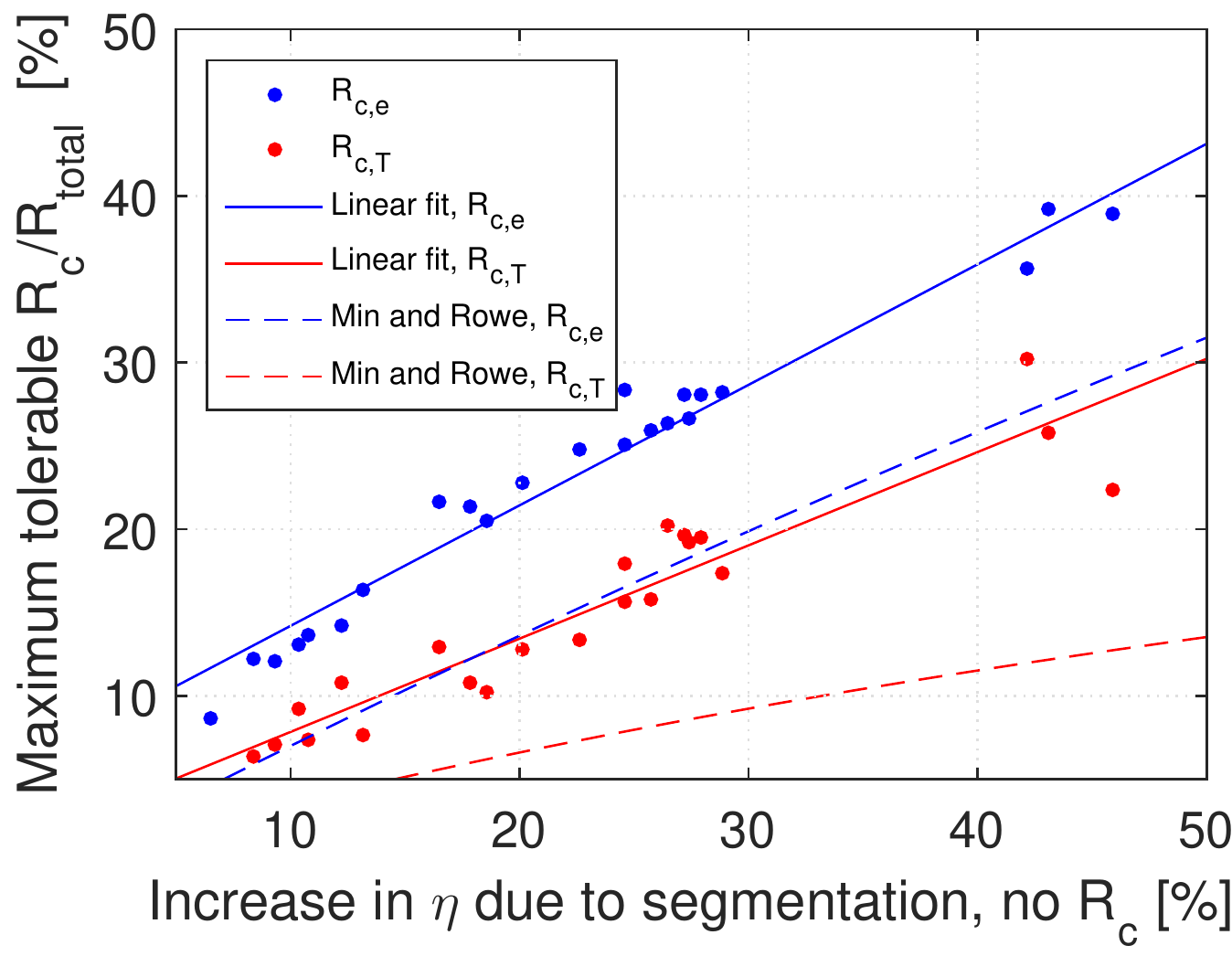}
  \caption{The maximum tolerable pure electrical or thermal contact resistance for a segmented TE leg as function of the gain in efficiency due to the segmentation. As the contour curves in Fig. \ref{Fig_Maximum_eff_seg_Rc} can be approximated as linear, the maximum tolerable combined electrical and thermal contact resistance can be found using a linear expression. The model of Min and Rowe \cite{Min_1992} is also shown.}
  \label{Fig_Performance_Rc}
\end{figure}

The individual curves in Fig. \ref{Fig_Maximum_eff_seg_Rc} can to a good approximation be considered linear. The maximum tolerable electrical contact resistance for the case of zero thermal contact resistance and vice versa is shown in Fig. \ref{Fig_Performance_Rc} for all systems. The points corresponds to the intersection of the lines in Fig. \ref{Fig_Maximum_eff_seg_Rc} with the axis. As can be seen from Fig. \ref{Fig_Performance_Rc} there is a clear linear behavior in both the electrical and thermal contact resistances. Thus if the total contact resistance is known for a given type of joining, one can use Fig. \ref{Fig_Performance_Rc} to determine if the segmented system will have an increased efficiency or not, compared to a leg of only the high temperature material. As an example, if segmentation increases the efficiency of 30\% then a electrical contact resistance of 30\% or a thermal contact resistance of 20\% can be tolerated.

The analytical expression given in Eq. (\ref{Eq.Min_Rowe}) is also shown in Fig. \ref{Fig_Performance_Rc}. As the analytical model assumes constant material properties, we assume that segmenting two legs results in an effective increase in $ZT$. Based on this, the maximum tolerable contact resistance can be found, as function of the increase in efficiency. This will depend on the unsegmented $ZT$ value of the material, which we here assume to be $ZT$ = 1. There is also a dependence of hot side temperature, but this is almost negligible for the contact resistance values considered here. As can be seen from the figure, for both the electrical and thermal contact resistance the analytical solution predicts a too low tolerance for contact resistance. This is similar to the trends observed in Figs. \ref{Fig_Rc_Eff} and \ref{Fig_Rt_Eff}.

Furthermore, since the individual curves are linear in Fig. \ref{Fig_Maximum_eff_seg_Rc}, the maximum tolerable electrical contact resistance can be predicted, for any known thermal contact resistance and vice versa. If the maximum tolerable electrical contact resistance for the case of zero thermal contact resistance, $R_\n{c,e,T=0}$, and the maximum tolerable thermal contact resistance for the case of zero electrical contact resistance, $R_\n{c,e=0,T}$, are known then the maximum tolerable electrical, $R_\n{c,e}$, or thermal, $R_\n{c,T}$, contact resistance can be calculated as
\begin{eqnarray}
R_\n{c,e} &=& -\frac{R_\n{c,e,T=0}}{R_\n{c,e=0,T}}R_\n{c,T}+R_\n{c,e,T=0}\nonumber\\
R_\n{c,T} &=& -\frac{R_\n{c,e=0,T}}{R_\n{c,e,T=0}}(R_\n{c,e}-R_\n{c,e,T=0})
\end{eqnarray}
assuming that the other contact resistance, i.e. the thermal or electrical contact resistance, is known.

\section{Conclusion}
The influence of electrical and thermal contact resistance on the efficiency of a segmented thermoelectric power generator was investigated. A total of 12 different segmented $p$-legs and 12 different segmented $n$-legs were investigated, using 8 different $p$-type and 8 $n$-type thermoelectric material. A universal influence of both the electrical and thermal contact resistance was observed for all system when the decrease in efficiency was examined as function of the fraction of contact resistance to the total resistance of the leg. The analytical model of Min and Rowe \cite{Min_1992} was shown not to accurately predict the decrease in efficiency as function of contact resistance for a segmented leg. In order for the efficiency not to decrease more than 20\%, the contact electrical resistance should be less than 30\% of the total leg resistance for zero thermal contact resistance, while the thermal contact resistance should be less than 20\% for zero electrical contact resistance. This universal behavior allowed the maximum tolerable contact resistance for a segmented system to be found, i.e. the resistance at which a leg of only the high temperature thermoelectric material has the same efficiency as the segmented system with a contact resistance at the interface. If e.g. segmentation increases the efficiency of 30\% then a electrical contact resistance of 30\% or a thermal contact resistance of 20\% can be tolerated.

\section*{Acknowledgements}
The author would like to thank the Programme Commission on Sustainable Energy and Environment, The Danish Council for Strategic Research for sponsoring the ``Oxide thermoelectrics for effective power generation from waste heat'' (OTE-POWER) (Project No. 10-093971) project as well as the ``CTEC - Center for Thermoelectric Energy Conversion'' (Project No. 1305-00002B) project. The author also wish to thank the European Commission for sponsoring the ``Nano-carbons for versatile power supply modules'' (NanoCaTe) (FP7-NMP Project No. 604647) project.

\end{document}